\documentclass[10pt,aps,fleqn,superscriptaddress,notitlepage,
  nofootinbib,preprintnumbers,nobalancelastpage]{revtex4-1}
\pdfoutput=1

\usepackage{pstricks,xspace}

\usepackage[latin1]{inputenc}

\usepackage{graphicx,overpic}

\usepackage{amsmath,amssymb,multirow,mathtools}

\usepackage{booktabs}

\usepackage{xifthen}

\usepackage[colorlinks=true, pdftex, linktocpage, citecolor=red, urlcolor=blue]{hyperref}

\newlength{\rivetLength}
\newcommand{\rivetFigure}[1]{
  \includegraphics[width=\rivetLength]{#1}
}
\newcommand{\rivetFigureRatio}[1]{
  \includegraphics[width=\rivetLength]{#1}\vspace*{-0.5mm}
}

\newcommand{\eqRef}[1]{eq.~\eqref{#1}\xspace}

\newcommand{\appRef}[1]{app.~\ref{#1}\xspace}

\newcommand{\figRef}[1]{fig.~\ref{#1}\xspace}

\newcommand{\FigsRef}[1]{Figs.~\ref{#1}\xspace}
\newcommand{\secRef}[1]{sec.~\ref{#1}\xspace}
\newcommand{\secsRef}[1]{secs.~\ref{#1}\xspace}

\newcommand{\me}{ME\xspace}

\newcommand{\mec}{MEC\xspace}
\newcommand{\mecs}{MECs\xspace}
\newcommand{\mops}{MOPS\xspace}
\newcommand{\mecd}{ME corrected\xspace}

\newcommand{\tsc}[1]{\textsc{#1}}
\newcommand{\tbf}[1]{\textbf{#1}}
\newcommand{\tit}[1]{\textit{#1}}
\newcommand{\mrm}[1]{\mathrm{#1}}

\renewcommand{\d}{\mrm{d}}
\renewcommand{\t}[1]{\mrm{#1}}

\newcommand{\scl}{t}
\newcommand{\scale}[1]{\ensuremath{\scl_{\,#1}}}
\newcommand{\facScale}{\scale{\t{fac}}}
\newcommand{\renScale}{\scale{\t{ren}}}
\newcommand{\cutScale}{\scale{\t{cut}}}
\newcommand{\startScale}{\scale{\t{start}}}
\newcommand{\maxScale}{\scale{\t{max}}}
\newcommand{\cutoff}{\mu_\mrm{\,c}}

\newcommand{\as}{\ensuremath{\alpha_s}\xspace}
\newcommand{\pimp}{\ensuremath{P_{\,\mrm{imp}}}}

\newcommand{\dSig}[2]{\d\sigma_{\,{#1}}(#2)}

\newcommand{\noEmi}[3]{\ensuremath{\Pi_{\,{#1}}({#2},{#3})}}
\newcommand{\noEmiUO}[3]{\ensuremath{\Pi^{\,\t{uo}}_{\,{#1}}({#2},{#3})}}
\newcommand{\sudakov}[3]{\ensuremath{\Delta_{\,{#1}}({#2},{#3})}}
\newcommand{\ME}[1]{\ensuremath{\left|\mc M{({\scriptstyle{#1}})}\right|^2}}
\newcommand{\MEC}[1]{\ensuremath{\mc R({\scriptstyle{#1}})}}
\newcommand{\pPhi}[1]{\ensuremath{\Phi_{\,{#1}}}}
\newcommand{\dPhi}[1]{\ensuremath{\t{d}\pPhi{#1}}}
\newcommand{\PDF}[3]{\ensuremath{f_{\,{#1}}(#2,#3)}}

\newcommand{\transZ}{\ensuremath{{\scriptstyle{\pPhi{0}}}}}
\newcommand{\transZP}{\ensuremath{{\scriptstyle{\pPhi{0}'}}}}

\newcommand{\transZtoO}{\ensuremath{{\scriptstyle{\pPhi{1}/\pPhi{0}}}}}
\newcommand{\transZPtoO}{\ensuremath{{\scriptstyle{\pPhi{1}/\pPhi{0}'}}}}
\newcommand{\transZPtoOP}{\ensuremath{{\scriptstyle{\pPhi{1}'/\pPhi{0}'}}}}

\newcommand{\transOPtoT}{\ensuremath{{\scriptstyle{\pPhi{2}/\pPhi{1}'}}}}
\newcommand{\transOPtoTP}{\ensuremath{{\scriptstyle{\pPhi{2}'/\pPhi{1}'}}}}

\newcommand{\transNtoNPO}{\ensuremath{{\scriptstyle{\pPhi{n+1}/\pPhi{n}}}}}
\newcommand{\transNPtoNPO}{\ensuremath{{\scriptstyle{\pPhi{n+1}/\pPhi{n}'}}}}

\newcommand{\transNMOPtoNP}{\ensuremath{{\scriptstyle{\pPhi{n}'/\pPhi{n-1}'}}}}

\newcommand{\transKPtoKPOP}{\ensuremath{{\scriptstyle{\pPhi{k+1}'/\pPhi{k}'}}}}
\newcommand{\transKPOPtoKPTP}{\ensuremath{{\scriptstyle{\pPhi{k+2}'/\pPhi{k+1}'}}}}

\newenvironment{newItem}
{\begin{list}{$\bullet$}{
  \setlength{\topsep}{2mm}\setlength{\partopsep}{2mm}
  \setlength{\itemsep}{1mm}\setlength{\parsep}{1mm}
}}
{\end{list}}
\newcounter{qArabic}
\newenvironment{enumArabic}
{\begin{list}{\tit{\arabic{qArabic})}}{
  \usecounter{qArabic}
  \setlength{\topsep}{1mm}\setlength{\partopsep}{1mm}
  \setlength{\itemsep}{1mm}\setlength{\parsep}{1mm}
}}
{\end{list}}
\newcounter{qAlph}
\newenvironment{enumAlph}
{\begin{list}{\tit{\alph{qAlph})}}{
  \usecounter{qAlph}
  \setlength{\topsep}{1mm}\setlength{\partopsep}{1mm}
  \setlength{\itemsep}{1mm}\setlength{\parsep}{1mm}
}}
{\end{list}}
\newcounter{qState}

\newlength{\abstwidth}
\newcommand{\Pythia}{P\protect\scalebox{0.8}{YTHIA}\xspace}

\newcommand{\Vincia}{V\protect\scalebox{0.8}{INCIA}\xspace}
\newcommand{\Madgraph}{MadGraph\xspace}
\newcommand{\lhapdf}{LHAPDF\xspace}

\newcommand{\mc}[1]{\mathcal{#1}}

\usepackage{hyperref}
\hypersetup{
  pdfauthor={Nadine Fischer, Stefan Prestel},
  pdftitle={Combining states without scale hierarchies with ordered parton showers},
  pdfkeywords={QCD, Parton Shower, NLO}
}

\begin{document}
\preprint{CoEPP-MN-17-4}
\preprint{FERMILAB-PUB-17-202-PPD-T}
\preprint{MCnet-17-09}
\title{Combining states without scale hierarchies with ordered parton showers}
\author{Nadine~Fischer}
\affiliation{School of Physics and Astronomy, Monash University, Clayton, VIC 3800, Australia}
\author{Stefan~Prestel}
\affiliation{Fermi National Accelerator Laboratory, Batavia, IL, 60510-0500, USA}
\begin{abstract}
We present a parameter-free scheme to combine fixed-order multi-jet results with
parton-shower evolution. The scheme produces jet cross sections with leading-order
accuracy in the complete phase space of multiple emissions, resumming large 
logarithms when appropriate, while not arbitrarily enforcing ordering on momentum
configurations beyond the reach of the parton-shower evolution equation. 
This requires the development of a matrix-element correction scheme for complex
phase-spaces including ordering conditions as well as a systematic scale-setting
procedure for unordered phase-space points. The resulting algorithm does not
require a merging-scale parameter. We implement the new
method in the \Vincia framework and compare to LHC data.
\end{abstract}

\maketitle

\graphicspath{{images/}}

\section{Introduction}
\label{sec:intro}

High-energy physics in the era of the Large Hadron Collider relies on accurate
calculations of Standard-Model scattering signatures -- both to determine 
backgrounds when directly searching for new physics and to allow for setting
indirect bounds by comparing measurements to precision calculations. Since
measurements at the LHC are typically sensitive to the formation and evolution
of jets, much attention has been devoted to calculating QCD corrections. This
has led to exquisite dedicated high-precision calculations, and to the
development of general schemes to overcome the limited applicability of 
individual fixed-order QCD calculations by combining multiple calculations into a 
single consistent result. To this end, modern General Purpose Event Generators
\cite{Buckley:2011ms,Gleisberg:2008ta,Sjostrand:2014zea,Bahr:2008pv} include 
a variety of complex matching~\cite{
  Frixione:2002ik,*Nason:2004rx,*Frixione:2007vw,*Frixione:2010ra,
  *Torrielli:2010aw,*Alioli:2010xd,*Hoeche:2010pf,*Hoeche:2011fd,
  *Platzer:2011bc,*Alwall:2014hca,*Jadach:2015mza,*Czakon:2015cla} 
 and merging~\cite{Catani:2001cc,*Mangano:2001xp,*Mrenna:2003if,*Alwall:2007fs,
  *Hamilton:2009ne,*Hamilton:2010wh,*Hoche:2010kg,*Lavesson:2008ah,
  *Lonnblad:2012ng, Lonnblad:2001iq,*Lavesson:2005xu,Lonnblad:2011xx,
  Platzer:2012bs,*Gehrmann:2012yg,*Hoeche:2012yf,*Lonnblad:2012ix,
  *Frederix:2012ps,*Alioli:2012fc,*Bellm:2017ktr}
schemes.

A unified Standard-Model prediction that is applicable for precision
measurements and new-physics searches alike must naturally include particle 
configurations that probe very different aspects of the calculation. The optimal
perturbative description of very different particle (and momentum) 
configurations can consequently vary significantly within one measurement, so
that care must be taken to avoid applying specialized arguments outside
of their region of validity. Otherwise, the accuracy of the calculation is
in jeopardy and its uncertainty might be underestimated. For example, applying 
QCD reasoning to events without large hierarchies in the hardness
of jets can lead to problematic effects \cite{Christiansen:2015jpa}.  

Standard-model calculations at the LHC can somewhat artificially be categorized
as focussing on momentum configurations with or without large
scale (hardness) hierarchies between jets.
Fixed-order QCD calculations are often appropriate for the latter, while
the former require a resummation of large perturbative enhancements by means
of evolution equations. Both approaches have complementary strengths and
should be combined for a state-of-the-art calculation. It is crucial to avoid
bias when constructing a single calculation that describes very different limits.
 
In this article, we design a new algorithm to combine multiple fixed-order
calculations for different parton multiplicities with each other and with
(parton-shower) resummation of large logarithmic enhancements. The aim of this
combined calculation is to simultaneously describe up to $n$ hard, well-separated 
partons with fixed-order matrix elements while retaining the jet evolution given
by the parton shower. We enforce strict 
requirements on the new scheme to improve on previous ideas:
\begin{enumerate}
\item The introduction of new parameters into the calculation is avoided. This
      is especially important when the correlation with existing parameters is
      not obvious.
\item The method should provide a uniform accuracy over the complete phase 
      space for one particle multiplicity. For now, this means that the rate
      of $n$ jets should be given with leading-order accuracy in QCD, 
      irrespectively of the hardness of jets.
\item The method should be largely agnostic to parton-shower-inspired arguments
      when configurations without large scale hierarchies are discussed.
\end{enumerate}
The resulting method borrows concepts from the CKKW-L method of merging 
matrix elements and parton showers~\cite{Lonnblad:2001iq,*Lavesson:2005xu,
Lonnblad:2011xx}, as well as 
from matrix-element correction schemes~\cite{Giele:2011cb,
Fischer:2016vfv}. We provide a new solution to the treatment of phase-space
regions beyond the reach of traditional shower evolution. Furthermore, we
improve upon the structure of the combined calculation in the parton-shower
region of soft and/or collinear emissions. Our new method consists of two 
main developments: the introduction (and implementation) of matrix-element
corrections for ordered parton-shower evolution, and the definition of a 
general scale-setting prescription based on matrix elements for 
contributions without apparent scale hierarchies. The benefit of using 
matrix-element corrections for shower-like splitting sequences is that unitarity
of fixed-order multi-jet cross sections is automatically guaranteed in these
phase-space regions. This means that the inclusive rates for $n$ jets will 
be correctly described with fixed order accuracy, without the need for explicit 
subtractions of negative weight, even if the rate for $n+1$ jets is also 
corrected with matrix elements. We will 
describe how the new method allows to achieve leading-order accuracy in
QCD for multi-parton configurations. This establishes a baseline for future 
developments beyond leading-order QCD. 

The new scheme relies on applying leading-order matrix-element corrections in 
phase-space regions that are accessible by a sequence of splittings ordered in 
a parton-shower evolution variable, supplemented with fixed-order results for 
configurations that cannot be reached by any such sequence.
We will use the misnomer \emph{``shower 
configurations"} for such states, and call states which cannot be reached by
an ordered sequence of shower emissions \emph{``non-shower states"}.

A very brief introduction to the parton-shower formalism and the notation is 
established in \secRef{sec:shower}.
The new method to iteratively correct parton showers with matrix elements
is described in detail in \secRef{sec:MOPS}. The combination of this scheme of
matrix-element corrections for ordered parton-shower evolution with non-shower 
states is discussed in \secRef{sec:nonShower}.
An executive summary of the algorithm is given in \secRef{sec:summary},
followed by a discussion of the impact of combining parton-shower-like 
and non-shower phase-space regions at parton level.
Then, results and data comparisons are presented in \secRef{sec:results} 
before we summarize and give an outlook in \secRef{sec:conclusions}. 
Additional details about the smoothly ordered showers and 
``GKS" matrix-element corrections
previously used in \Vincia are collected in \appRef{app:GKSMECs}, while a 
thorough validation of new matrix-element corrections for ordered parton-shower 
evolution is given in \appRef{app:validation}.

\section{Parton showers and matrix element corrections}
\label{sec:shower}

To set the scene and establish notation, let us briefly review some parton-shower 
basics. We start by defining the effect of parton-shower 
evolution~\cite{Sjostrand:1985xi,Marchesini:1987cf}
on an arbitrary observable $O$ (in the notation of~\cite{Hoche:2015sya}),
\begin{align}\label{eq:shower_functional}
  \mc{F}_{\vec{a}}(\Phi_n,t,t';O)=\;\mc{F}_{\vec{a}}(\Phi_n,t,t')\,O(\Phi_n)
  +\int_t^{t'}\frac{\d\bar{t}}{\bar{t}}\,
  \frac{\d\mc{F}_{\vec{a}}(\Phi_n,\bar{t},t')}{\d\ln\bar{t}}\,
  \mc{F}_{\vec{a}'}(\Phi'_{n+1},t,\bar{t};O)~,
\end{align}
where  $t\equiv t(\transNtoNPO)$ is the shower evolution variable, and the 
shower generating functional $\mc{F}$ depends on the list of parton flavors 
$\vec{a}$, and the corresponding $n$-particle momentum configuration $\Phi_n$. 
Though not explicitly stated, any $n$-particle state contains an arbitrarily 
complicated Born state, $\pPhi{n}\equiv \pPhi{\t{B}+n}$.
The first term in \eqRef{eq:shower_functional} encodes the contribution from
no resolvable shower emissions, while the second piece includes one or more
emissions. The parton flavors $\vec{a}'$ of the $(n+1)$-particle momentum 
configuration $\Phi_{n+1}$ include the resolved emission and the partons 
$\vec{a}$, with momenta changed according to the recoil prescription of the 
parton shower and flavor changes where applicable.
The generating functional obeys the evolution equation
\begin{align}
\label{eq:PSevol}
  \frac{\d\ln\mc{F}_{\vec{a}}(\Phi_n,t,\mu^2)}{\d t}
  =\sum_{i\in{\rm IS}}\sum_{b=q,g}\int_{x_i}^{1-\varepsilon}\frac{\d z}{z}\,
  \frac{\alpha_s(t)}{2\pi}\,P_{ba_i}\,
  \frac{f_{b}(x_i/z,t)}{f_{a_i}(x_i,t)}
  +\sum_{j\in{\rm FS}}\sum_{b=q,g}\int_\varepsilon^{1-\varepsilon}\d z\,
  \frac{\alpha_s(t)}{2\pi}\,P_{a_jb}~,
\end{align}
where $z\equiv z(\transNtoNPO)$ is an energy-sharing variable and $x$ the 
momentum fraction of the incoming parton in $\Phi_n$. The first term in 
\eqRef{eq:PSevol} corresponds to evolution by initial-state radiation, while 
the second term represents final-state radiation. Backward 
evolution~\cite{Sjostrand:1985xi} for initial-state radiation introduces a 
ratio of parton distribution functions (PDFs) $f$ in the first term. The 
quality of the shower real-radiation pattern is governed by the unregularized, 
dimensionful splitting kernels $P_{ij}\equiv P_{ij}(\transNtoNPO)$\,\footnote{
We define $P_{ij}(\transNtoNPO)$ as dimensionful to follow the convention
used in the antenna literature~\cite{Kosower:1997zr,*GehrmannDeRidder:2005cm}. 
Thus, $P_{ij}$ corresponds to 
$\hat P_{ij}/t$ in the notation of~\cite{Hoche:2015sya}, leading to a 
marginally different notation compared to the latter.}.
For brevity, we will suppress the indices
of the splitting functions. The shower will produce an accurate real-emission 
pattern if the sum of all products of splitting probabilities and transition 
probabilities $\ME{\pPhi{n}}$ is a good approximation of the full real-emission 
probability $\ME{\pPhi{n+1}}$. For a transition from an $n$-particle to an
$(n+1)$-particle state, this can be achieved by the (symbolic) replacement
\begin{align}
  & \left[\sum_{\pPhi{n}} P(\transNtoNPO)\ME{\pPhi{n}} \right] ~~
  \rightarrow \nonumber \\
  & \left[\sum_{\pPhi{n}} P(\transNtoNPO)\ME{\pPhi{n}} \right] 
  \frac{\ME{\pPhi{n+1}}}
  {\left(\sum_{\pPhi{n}'} P(\transNPtoNPO)\ME{\pPhi{n}'} \right)}
  = \sum_{\pPhi{n}} \left[P(\transNtoNPO)\ME{\pPhi{n}} \MEC{\pPhi{n+1}} \right]
  ~.
\end{align}
Such a process- and multiplicity-dependent redefinition of the splitting
kernel is called matrix-element correction (\mec). It is worth noting that this
replacement changes both the shower no-emission probability and the
real-emission pattern. The real-emission pattern is corrected to a target
fixed-order accuracy. However, the accuracy of the parton-shower resummation of
virtual corrections into Sudakov factors is not improved.

The impact of \me corrections is largest for hard, well-separated jets, as 
splitting kernels do not approximate the full fixed-order matrix element well 
for configurations with hard, well-separated jets. Thus, the most significant
improvement of \me corrections can be obtained when correcting the 
$n$ hardest splittings
in the shower cascade. In practise, this means that hardness-ordered parton
showers allow for simpler \mec schemes~\cite{Bengtsson:1986hr,*Bengtsson:1986et,*Miu:1998ju}, which in particular do not require 
knowledge of high-multiplicity matrix elements as a function of evolution 
variables\,\footnote{A scheme to correct the hardest emission in angular-ordered 
showers has been discussed in \cite{Seymour:1994df}. This scheme requires to
apply the same correction repeatedly, to guarantee that the single hardest
emission is corrected to leading-order accuracy. Although promising from the
resummation standpoint, it is, however, not obvious how this scheme could be used
to correct the $n$ hardest emissions.}. Instead, it is sufficient that the 
parton shower generates complete,
physical intermediate momenta $\pPhi{n}$ that can be used to evaluate
$\ME{\pPhi{n}}$ numerically. Thus, we will limit our discussion to 
hardness-ordered shower programs. This will allow for a level of process-independence, 
and make the \emph{iteration} of \me corrections possible.

The key technical difficulty for a consistent application of \me 
corrections is the construction of 
the sum over parton-shower paths in the denominator of the correction factor
$\MEC{\pPhi{n+1}}$. Since parton showers are formulated as
Markov processes, neither the weight nor the existence of each term in the 
sum is known a priori when the splitting governed by $P(\transNtoNPO)$ is generated, and all terms
have to be reconstructed explicitly.

\section{Matrix-element corrections for ordered parton showers}
\label{sec:MOPS}

The formalism of \me corrections for ordered parton showers
(\mops) is close in spirit to the idea of the iterative \mec approach
of \cite{Giele:2011cb,Fischer:2016vfv}\,\footnote{A short review of the GKS approach is given in \appRef{app:GKSMECs}.}. These previous ideas rely on a history-independent 
parton shower that is able to fill the complete available phase space. 
This necessitates abandoning parton-shower ordering, i.e.\ the property that
ensures the resummation of large logarithms in ratios of evolution scales. 
Sensible resummation properties then rely on the introduction of auxiliary 
functions. Furthermore, configurations with hard well-separated jets might 
contain poorly understood higher-order corrections. It is thus sensible to
limit \me corrections for the parton shower to phase-space regions
reachable by an ordered sequence of branchings. This means that we need to
rederive appropriate MEC factors $\MEC{\pPhi{n}}$ that correctly encode
the presence of complicated phase-space constraints due to ordering -- making
the resulting method substantially different from previous attempts. 

To not overcomplicate the derivation of the \mops formalism, we drop all 
coupling- and PDF factors in this section. These pieces are evaluated exactly
as in an uncorrected parton shower (the probability of a splitting at
evolution scale $t$ includes a factor $\alpha_s(t)/2\pi$, splittings involving 
initial legs induce ratios of PDFs $f(\frac{x}{z},t)/f(x,t)$, cf.~\eqRef{eq:PSevol}), and do not enter 
in the \mec factors. Similarly, Sudakov factors are not explicitly 
written out when demonstrating the \mops method. The \mops 
procedure is applied during the Sudakov veto-algorithm as a redefinition of the splitting
kernels, meaning that both the (real) emission probability and the 
no-emission probabilities are \mecd. This ensures the unitarity of the method, 
i.e.~that corrections to higher parton multiplicities vanish in observables
that are only sensitive to a lower multiplicity.

Consider an arbitrary Born process with factorization scale $\facScale\equiv
\scl(\transZ)$ as starting point of the parton shower. The weight of the first
branching is
\begin{align}\label{eq:shower1}
  P(\transZtoO)~\Theta(\scl(\transZ)-\scl(\transZtoO))~
  \ME{\pPhi{0}}~\dPhi{1}~,
\end{align}
where the shower is restricted to scales below the factorization scale.
For processes that require regularizing cuts at Born level, the matrix element
$\ME{\pPhi{0}}$
can be suitable redefined to include the necessary $\Theta$-functions.
To correct the weight of the phase-space point $\pPhi{1}$ to the full
fixed-order matrix element, all possible emissions from ``underlying" 
Born configurations 
$\pPhi{0}'$ that could have produced the phase-space point 
$\pPhi{1}$ that we want to correct have to be taken into account.
A suitable multiplicative correction factor is thus
\begin{align}\label{eq:MECfactor1}
  \MEC{\pPhi{1}} = \frac{\ME{\pPhi{1}}}
  {\sum\limits_{\pPhi{0}'} P(\transZPtoO)~
  \Theta(\scl(\transZP)-\scl(\transZPtoO))~
  \ME{\pPhi{0}'} }~.
\end{align}
Applying this correction to each individual splitting and summing over all
shower contributions cancels the denominator of \eqRef{eq:MECfactor1} 
and gives
\begin{align}
  \MEC{\pPhi{1}}~\sum_{\pPhi{0}} 
  P(\transZtoO)~\Theta(\scl(\transZ)-\scl(\transZtoO))~
  \ME{\pPhi{0}} = \ME{\pPhi{1}}~.
\end{align}
The calculation of the correction factor for the weight of a second branching
becomes more cumbersome,
\begin{flalign}\label{eq:MECfactor2}
  \MEC{\pPhi{2}} = \frac{\ME{\pPhi{2}}}
  {\sum\limits_{\pPhi{1}'}P(\transOPtoT)~\MEC{\pPhi{1}'}
  \sum\limits_{\pPhi{0}'}\Theta(\scl(\transZPtoOP)-\scl(\transOPtoT))~
  P(\transZPtoOP)~\Theta(\scl(\transZP)-\scl(\transZPtoOP))
  \ME{\pPhi{0}'}}~.
\end{flalign}
Here, the denominator sums over all possible ways how the shower can populate
the phase-space point $\pPhi{2}$, taking into account the allowed (ordered)
paths through the $\Theta$-functions with the \mecd parton-shower weights
of the intermediate $+1$-particle phase-space points. 
Consequently, $\MEC{\pPhi{2}}$ includes the correction factors
of the previous order, $\MEC{\pPhi{1}'}$.

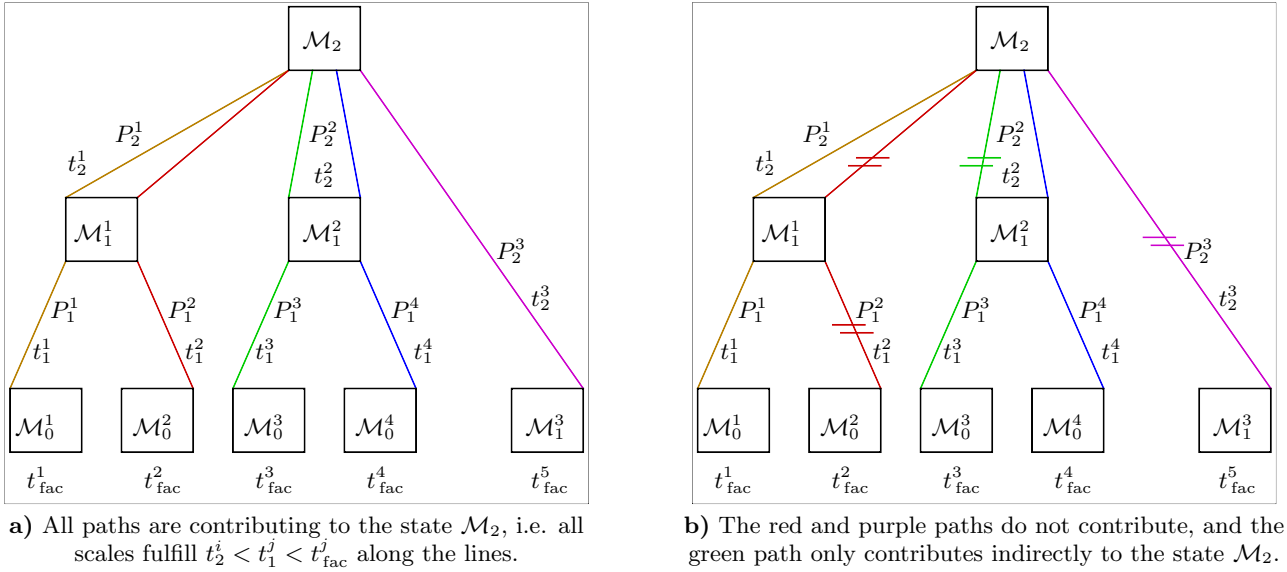
\begin{figure}[t]
\centering
\begin{minipage}[t]{0.49\textwidth}
\begin{overpic}[scale=0.5]{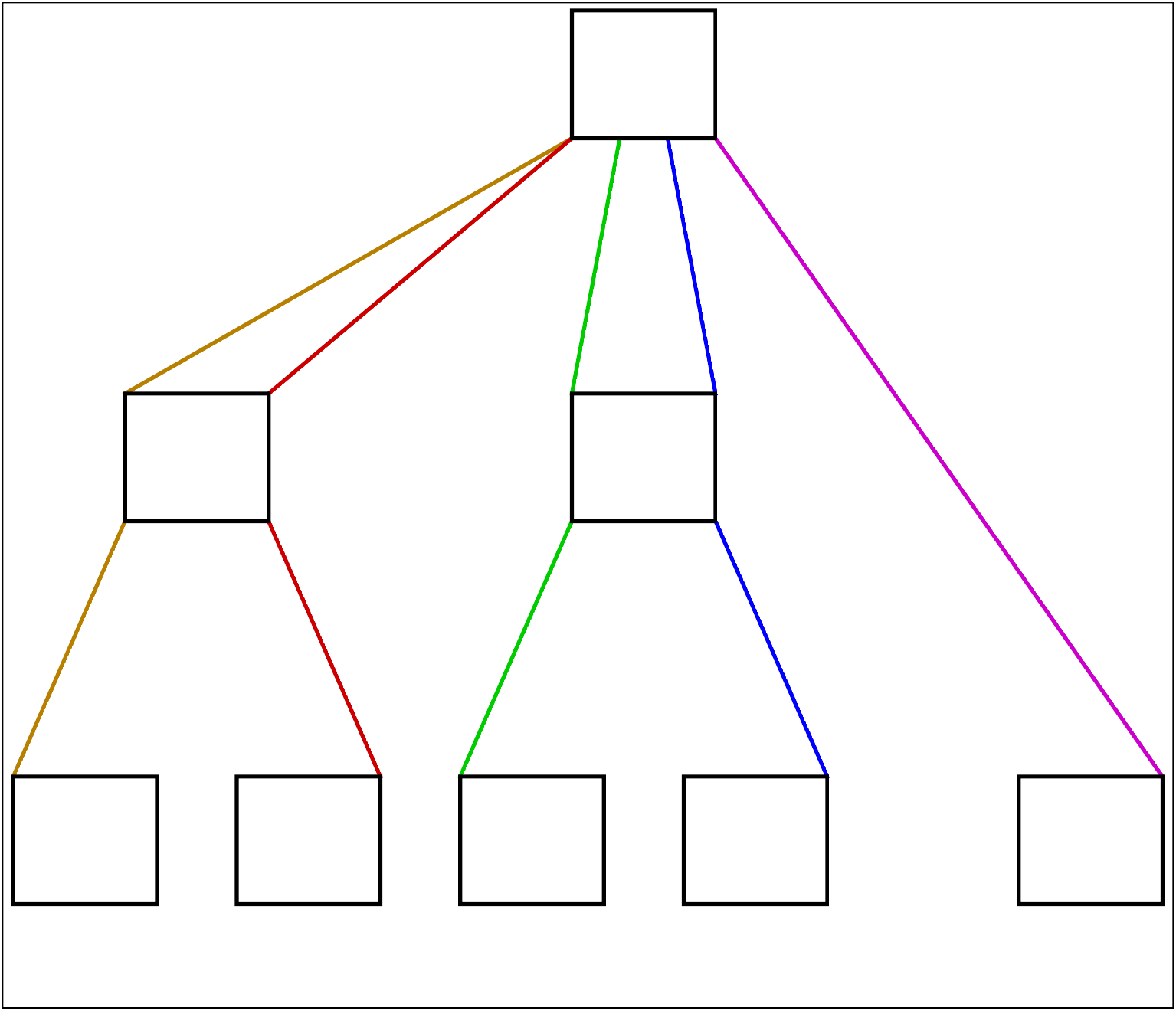}
  \put( 4, 3){$\facScale^1$} \put(24, 3){$\facScale^2$}
  \put(43, 3){$\facScale^3$} \put(62, 3){$\facScale^4$}
  \put(90, 3){$\facScale^5$}
  \put( 2,12){$\mc M_0^1$} \put(22,12){$\mc M_0^2$} 
  \put(41,12){$\mc M_0^3$} \put(60,12){$\mc M_0^4$} \put(89,12){$\mc M_1^3$}
  \put( 5,25){$\scl_1^1$} \put( 8,32){$P_1^1$}
  \put(31,25){$\scl_1^2$} \put(28,32){$P_1^2$}
  \put(43,25){$\scl_1^3$} \put(46,32){$P_1^3$}
  \put(70,25){$\scl_1^4$} \put(66,32){$P_1^4$}
  \put(12,45){$\mc M_1^1$} \put(51,45){$\mc M_1^2$}
  \put(51,78){$\mc M_2$}
  \put(11,57){$\scl_2^1$} \put(19,62){$P_2^1$}
  \put(53,55){$\scl_2^2$} \put(52,62){$P_2^2$}
  \put(90,34){$\scl_2^3$} \put(84,42){$P_2^3$}
\end{overpic}
\\
\tbf{a)} All paths are contributing to the state $\mc M_2$, i.e. all \\
scales fulfill $\scl_2^i<\scl_1^j<\facScale^j$ along the lines.
\end{minipage}
\hfill
\begin{minipage}[t]{0.49\textwidth}
\begin{overpic}[scale=0.5]{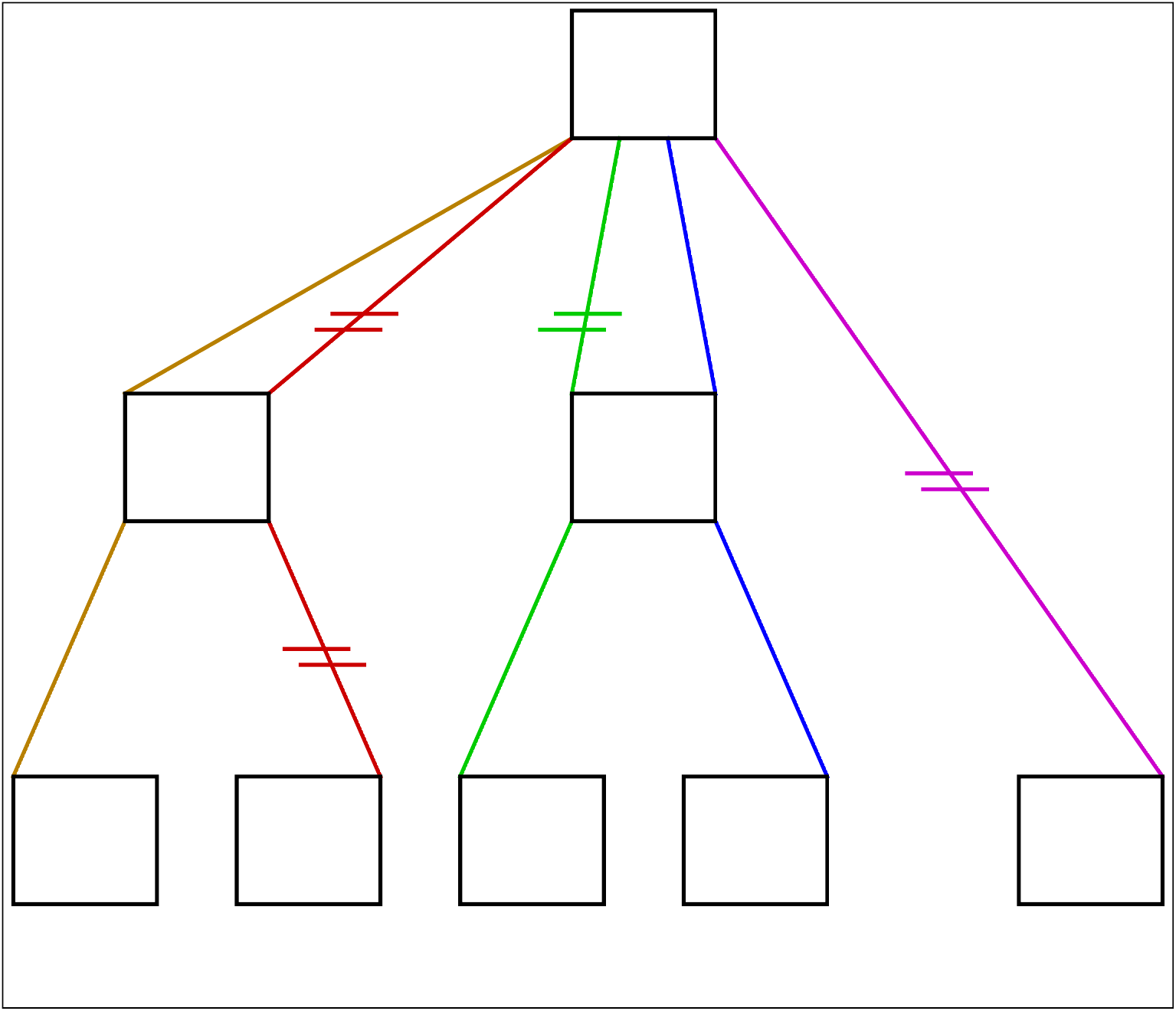}
  \put( 4, 3){$\facScale^1$} \put(24, 3){$\facScale^2$}
  \put(43, 3){$\facScale^3$} \put(62, 3){$\facScale^4$}
  \put(90, 3){$\facScale^5$}
  \put( 2,12){$\mc M_0^1$} \put(22,12){$\mc M_0^2$} 
  \put(41,12){$\mc M_0^3$} \put(60,12){$\mc M_0^4$} \put(89,12){$\mc M_1^3$}
  \put( 5,25){$\scl_1^1$} \put( 8,32){$P_1^1$}
  \put(31,25){$\scl_1^2$} \put(28,32){$P_1^2$}
  \put(43,25){$\scl_1^3$} \put(46,32){$P_1^3$}
  \put(70,25){$\scl_1^4$} \put(66,32){$P_1^4$}
  \put(12,45){$\mc M_1^1$} \put(51,45){$\mc M_1^2$}
  \put(51,78){$\mc M_2$}
  \put(11,57){$\scl_2^1$} \put(19,62){$P_2^1$}
  \put(53,55){$\scl_2^2$} \put(52,62){$P_2^2$}
  \put(90,34){$\scl_2^3$} \put(84,42){$P_2^3$}
\end{overpic}
\\
\tbf{b)} The red and purple paths do not contribute, and the \\
green path only contributes indirectly to the state $\mc M_2$.
\end{minipage}
\caption{History pyramid to illustrate different levels of contribution to the
\mops\ factor. The superscripts are numbering the different nodes. We use the 
shorthands $\mc M_{X}\equiv\ME{\pPhi{X}}$,
$\scl_X\equiv\scl({\scriptstyle{\pPhi{X}/\pPhi{X-1}}})$, and
$P_X\equiv P({\scriptstyle{\pPhi{X}/\pPhi{X-1}}})$.
The top layer is the main $+2$-particle state and the lower boxes 
represent the clustered states after one and two successive clusterings 
respectively. The scales and splitting probabilities associated with the 
clusterings are noted along the lines.
For illustrative purposes we included a path where the Born state is 
reached after one clustering (purple line), as present e.g. when combining 
QCD and electroweak clusterings.
\label{fig:paths}}
\end{figure}

It is useful to illustrate how this relatively complicated recursive
definition is obtained with an example. Consider the case of a $+2$-particle state
shown in \figRef{fig:paths}. The $+2$-particle state on top of
the pyramid can be reached from the base of the pyramid by several splitting 
sequences or ``paths". The paths are not 
necessarily physical but rather serve the purpose of illustration.
In \figRef{fig:paths} \tbf{a)} all paths directly contribute to the 
$+2$-particle
state, as each path from the base to the top follows a decreasing (i.e.\ ordered)
sequence of branchings scales. With the 
shorthands introduced in the caption of \figRef{fig:paths} the 
correction factors for the $+1$-particle states are
\begin{align}
  \mc R_1^1 = \frac{\mc M_1^1}{P_1^1\,\mc M_0^1 + P_1^2\,\mc M_0^2}
  \quad\text{and}\quad
  \mc R_1^2 = \frac{\mc M_1^2}{P_1^3\,\mc M_0^3 + P_1^4\,\mc M_0^4}~.
\end{align}
Both factors contribute to the correction to the $+2$-particle state,
\begin{align}\label{eq:R2example1}
  \mc R_2 = \frac{\mc M_2}{P_2^1\,\mc R_1^1\,(P_1^1\,\mc M_0^1 + 
  P_1^2\,\mc M_0^2) + P_2^2\,\mc R_1^2\,(P_1^3\,\mc M_0^3 + P_1^4\,\mc M_0^4)
  + P_2^3\,\mc M_1^3} 
  = \frac{\mc M_2}{P_2^1\,\mc M_1^1+P_2^2\,\mc M_1^2+P_2^3\,\mc M_1^3}~.
\end{align}
Since all paths contribute, the nesting of the \mops factors cancels and the 
denominator reduces to the sum of the splitting kernels, multiplied with the 
$+1$-particle matrix elements.

Some paths in \figRef{fig:paths} \tbf{b)} are unordered, which leads to
$+1$ \mops factors of
\begin{align}
  \mc R_1^1 = \frac{\mc M_1^1}{P_1^1\,\mc M_0^1}
  \quad\text{and}\quad
  \mc R_1^2 = \frac{\mc M_1^2}{P_1^3\,\mc M_0^3 + P_1^4\,\mc M_0^4}~.
\end{align}
Only one path (brown) contributes to the denominator of $\mc R_1^1$ --
the other path (red) is unordered.

The correction to the $+2$-particle state is
\begin{align}\label{eq:R2example2}
  \mc R_2 = \frac{\mc M_2}{P_2^1\,\mc R_1^1\,P_1^1\,\mc M_0^1 +
  P_2^2\,\mc R_1^2\,P_1^4\,\mc M_0^4}
  = \frac{\mc M_2}{P_2^1\,\mc M_1^1
  + P_2^2\,\dfrac{\mc M_1^2}{P_1^3\,\mc M_0^3 + P_1^4\,\mc M_0^4}
  \,P_1^4\,\mc M_0^4}~.
\end{align}
The red path in \figRef{fig:paths} \tbf{b)} does not contribute at all to the 
$+2$-particle state since the first branching scale is exceeding the 
factorization scale, $\scl_1^2>\facScale^2$. This leads to a cancellation in 
the first term of the denominator in \eqRef{eq:R2example2}. The green path is 
not contributing directly to the $+2$-particle state, as $\scl_2^2>\scl_1^3$. 
However, since $\scl_1^3<\facScale^3$, the path is present in $\mc R_1^2$. 
Therefore, the \mops factor for the $+2$-particle
state depends on more than one ``layer" in the paths, and can thus not be
calculated by reconstructing only $+1$-particle states from the 
$+2$-particle state that should be corrected. 

This example highlights the core features of the \mops method.
The recursive structure of the correction factor represents a crucial 
difference to the GKS method (see \appRef{app:GKSMECs}). At first sight, 
it seems counter-intuitive that knowledge of ordered \emph{and} unordered 
paths is required to correctly calculate the correction factor for a 
phase-space point that has been generated by an ordered sequence of splittings.
However, the necessity becomes clear when the weight of intermediate
states is taken into account.

To obtain a uniform accuracy over the complete $n$-parton phase space,
states beyond the reach of the parton shower have to be included.
We discuss the treatment of these non-shower states in the next section, and 
present the general formula for the \mops factor in \secRef{sec:summary}.

\section{Completing the calculation with non-shower configurations}
\label{sec:nonShower}

The \mops formalism discussed in \secRef{sec:MOPS} only covers the
parton-shower phase space characterized by an ordered sequence of splitting
scales ($\facScale>t_1>t_2\dots$). As a 
consequence, a prescription for the missing phase space is required. The precise 
definition of these regions depends on the parton shower itself, the starting 
scale, definition of the evolution variable, and recoil strategy. 
Configurations can either be forbidden by restricting the first emission 
to scales below the factorization scale, cuts on lowest-multiplicity 
phase space points, or by the ordering property of the shower.

When combining non-shower and shower states, care has to be taken to avoid
double- or under-counting. As discussed in \secRef{sec:MOPS}, the shower off
lowest-multiplicity events is treated without any restrictions apart from 
ordering emissions in the parton-shower evolution variable. 
Only those higher-multiplicity states that cannot be reproduced by
showering lower-multiplicity states need to be added explicitly. This criterion 
supersedes algorithms that rely on the introduction 
of a merging cut\,\footnote{An arbitrary shower will not correctly 
describe \emph{all} sub-leading logs in its evolution variable, so that 
non-shower configurations may still contain (sub-leading) logarithmic 
divergences. One famous example of such configurations are the unordered, balancing
soft-gluon emissions leading to Parisi-Petronzio scaling in $p_\perp$ 
distributions \cite{Parisi:1979se,Frixione:1998dw}. To avoid such divergences in practice, we only include 
non-shower phase-space points
if each scale at which partons could be recombined (as defined by the 
shower evolution variable) is above the parton-shower cut-off 
$\cutScale\approx 1\textnormal{GeV}$.}. Uniform
(leading-order) accuracy then is obtained across the complete emission phase 
space by also applying a \mecd shower 
when adding soft-collinear shower radiation to non-shower states.
This will, if performed naively, introduce overlap between (the shower off) 
different non-shower states. Three steps are required to avoid the overlap:
\begin{enumArabic}
\item Non-shower events are defined as unordered if no ordered path exists, i.e. if
different paths to the same \me state are present, the event is only considered 
unordered if none of the paths can be reproduced with an ordered sequence of 
branchings scales.
\item Potential overlap between non-shower states with different parton
multiplicities has to be removed, e.g. a maximally unordered 
$+2$-particle state may also be produced as a shower emission off a 
maximally unordered $+1$-particle state. The explanation
how this overlap is identified and removed in the higher-multiplicity states,
is deferred to the end of~\secRef{sec:summary}, since it is helpful to first
discuss how non-shower states are showered.
\item States produced by ordered parton showers overlap with 
soft-collinear radiation attached to non-shower states if the ``history" of a
phase-space point contains both ordered and unordered paths.
Therefore, both have to be \mecd with correction factors taking into account
both possibilities of population.
\end{enumArabic}

We now turn to the scale setting in non-shower events with two or more 
additional partons. From a parton-shower standpoint, there is no a priori guideline how non-shower
configurations should be treated. However, since non-shower configurations 
easily dominate LHC observables depending on many well-separated jets, finding 
a sensible scale-setting prescription for arbitrary processes will
greatly improve the ability of fixed-order + parton-shower calculations to 
describe data. Variations around the central scale can then be used to assess
the precision of the calculation.   

To obtain a flexible scale-setting prescription, we borrow the idea of
constructing all possible event histories from the CKKW-L~\cite{Lonnblad:2001iq,*Lavesson:2005xu,Lonnblad:2011xx}
The aim of
the procedure is twofold: define dynamical scales by exploiting the information
about the phase-space points with the help of the weight and 
``substructure" of multi-jet matrix elements, while further ensuring a
smooth inclusion of non-shower states with shower-accessible events.

For a sensible scale-setting prescription for non-shower states,
we follow an argument similar to the derivation of the \mops factor.
However, ordering considerations should not be applied to non-shower
states. Assume that a phase-space point $\pPhi{n+1}$ can be reached from
multiple $\pPhi{n}'$ states by splitting an external leg. The contribution to
the cross section due to splitting a single leg can be approximated by
\begin{align}
  \as(\scl({\scriptstyle{\pPhi{n+1}/\pPhi{n}'}}))~
  P({\scriptstyle{\pPhi{n+1}/\pPhi{n}'}})~
  \as^{n}(\scale{n}^\mrm{\,eff})~\ME{\pPhi{n}'}~,
\end{align} 
where $\scale{n}^\mrm{\,eff}$ is a suitable scale for the 
``underlying" $n$-particle state. To obtain the correct (leading-order)
result when summing over all possible splittings 
$\pPhi{n}'\rightarrow \pPhi{n+1}$, we can apply the corrective factor
\begin{align}\label{eq:strongMECsSimple}
  \MEC{\pPhi{n+1}} = 
  \frac{\as^{n+1}(\scale{n+1}^\mrm{\,eff})~
  \ME{\pPhi{n+1}}}{\sum\limits_{\pPhi{n}'}
  \as(\scl({\scriptstyle{\pPhi{n+1}/\pPhi{n}'}}))~
  P({\scriptstyle{\pPhi{n+1}/\pPhi{n}'}})~
  \as^{n}(\scale{n}^\mrm{\,eff})~\ME{\pPhi{n}'}}~,
\end{align}
where $\scale{n+1}^\mrm{\,eff}$ is the desired (currently unknown) scale for the 
$(n+1)$-particle state.
To find a suitable scale, note that
\begin{enumAlph}
\item if one splitting dominates over all  
other splittings, then a natural scale to capture the dynamics is strongly 
correlated with the relative jet separation of the dominant splitting,
\item if no splitting dominates, i.e. all splittings contribute 
democratically, there should be no strong preference for a scale, and a 
weighted average of jet separations seems appropriate.
\end{enumAlph} 
Leaving aside the complications (and bias) induced by ordering constraints,
an identical argument holds for parton-shower-produced states. In this case,
the requirements above are fulfilled by keeping the characteristic shower-induced
$\alpha_s$ factors for every \mecd shower splitting. This would be
guaranteed if the $\as$ factors in \eqRef{eq:strongMECsSimple} would be identified
by
\begin{align}\label{eq:strongMECsAS}
  \as^{n+1}(\scale{n+1}^\mrm{\,eff}) = 
  \frac{\sum\limits_{\pPhi{n}'}
  \as(\scl({\scriptstyle{\pPhi{n+1}/\pPhi{n}'}}))~
  P({\scriptstyle{\pPhi{n+1}/\pPhi{n}'}})~
  \as^{n}(\scale{n}^\mrm{\,eff})~
  \ME{\pPhi{n}'}} {\sum\limits_{\pPhi{n}'}
  P({\scriptstyle{\pPhi{n+1}/\pPhi{n}'}})~\ME
  {\pPhi{n}'}}~,
\end{align}
since then, \eqRef{eq:strongMECsSimple} is a simplified MEC factor.
For ordered parton-shower sequences, \eqRef{eq:strongMECsAS} will not lead to 
the correct result. It is, however, well-suited as a scale-setting prescription 
for non-shower configurations. We will use \eqRef{eq:strongMECsAS} as the 
definition of the effective scales 
below, i.e.\ we set the renormalization and factorization scales for non-shower 
events to $\scl^\mrm{\,eff}$. The effective scale also serves as a shower 
(re)starting scale. The variation of the 
effective scale may act as an uncertainty estimate of the prescription.

An expression for the effective scale could also have been obtained by including
PDF ratios in \eqRef{eq:strongMECsSimple}, which would mean that the choice
of effective scale captured dynamics of underlying ``hadronic" cross sections.
We do not implement such a scale-setting prescription since we believe that the 
scale setting should be based on perturbative parton-level quantities.

Note that the scale-setting mechanism in \eqRef{eq:strongMECsAS} allows for
$\facScale < \scl^\mrm{\,eff}$ if the scales entering the calculation are 
sufficiently large. An example of such a configuration are non-shower 
states with multiple hard (and possibly balancing) jets without $p_\perp$ 
hierarchy. In this case, using a scale defined for the lowest-multiplicity
process can result in pathologies \cite{Berger:2009ep}. It is desirable that 
$\scl^\mrm{\,eff}$ is not bounded by \facScale, the 
factorization scale assigned to a fictitious lowest-multiplicity process. 
Instead, $\scl^\mrm{\,eff}$ should provide a more ``natural" scale for this
genuine multi-jet configuration. Furthermore, $\scl^\mrm{\,eff}$ is bound to
remain in the perturbative region, since we only include non-shower phase 
space points for which clustering scales (as defined by the 
shower evolution variable) are above the parton-shower cut-off.

In \secRef{sec:results} we will show that the scale setting outlined in this 
section results in a very good description of LHC data.

\section{The complete algorithm}
\label{sec:summary}

In this section, we summarize the combined fixed-order + parton-shower 
algorithm, and present the general form of the \mops factor.
The scheme introduces \me correction for several ordered consecutive 
parton-shower emissions. This is in general obtained by applying the
\mops factor 
\begin{align}
  \MEC{\pPhi{n+1}} = \ME{\pPhi{n+1}}\Bigg[
  &~\sum\limits_{\pPhi{n}'} P(\transNPtoNPO)~\MEC{\pPhi{n}'}~
  \sum\limits_{\pPhi{n-1}'} \Theta(\scl(\transNMOPtoNP)-
  \scl(\transNPtoNPO))~P(\transNMOPtoNP)~\MEC{\pPhi{n-1}'} \nonumber \\
  &~\prod\limits_{k=n-2}^{k\le1} \left( \sum\limits_{\pPhi{k}'}
  \Theta(\scl(\transKPtoKPOP)-\scl(\transKPOPtoKPTP))~
  P(\transKPtoKPOP)~\MEC{\pPhi{k}}\right)
  \nonumber \\[-1mm]
  &~\sum\limits_{\pPhi{0}'}
  \Theta(\scl(\transZPtoOP)-\scl(\transOPtoTP))~P(\transZPtoOP)~
  \Theta(\scl(\transZP)-\scl(\transZPtoOP))~\ME{\pPhi{0}'}
  \Bigg]^{-1}
  \label{eq:strongMECs}
\end{align}
to the splitting kernel. When including the correct weight of each possible 
path, the result exhibits a recursive structure, where $\MEC{\pPhi{n+1}}$
includes the correction factors of all previous orders, $\MEC{\pPhi{n}'}$ to 
$\MEC{\pPhi{1}'}$.
Once non-shower states
are added, their contributions to the \mops factor are taken into account 
as well.

Non-shower states are added as new configurations, with renormalization and 
factorization scales calculated through
\begin{align}\label{eq:teff}
  \as^{n+1}(\scale{n+1}^\mrm{\,eff}) = 
  \frac{\sum\limits_{\pPhi{n}'}
  \as(\scl({\scriptstyle{\pPhi{n+1}/\pPhi{n}'}}))~
  P({\scriptstyle{\pPhi{n+1}/\pPhi{n}'}})~
  \as^{n}(\scale{n}^\mrm{\,eff})~
  \ME{\pPhi{n}'}} {\sum\limits_{\pPhi{n}'}
  P({\scriptstyle{\pPhi{n+1}/\pPhi{n}'}})~\ME
  {\pPhi{n}'}}~.
\end{align}
This should ensure that the dynamics of the process are encoded in a 
sensible scale choice, without the scale-setting prescription being based on 
process- or multiplicity-dependent arguments.

Since non-shower states are
included without a hard cut-off (e.g.~a merging scale), the effective scale
$\scl^\mrm{\,eff}$ may differ significantly from the factorization scale
$\facScale$. In this case, we further attach Sudakov factors by means of 
trial showering~\cite{Lonnblad:2001iq}
to the non-shower states to include a 
sensible suppression due to the resummation of large logarithms 
of $\facScale/\scl^\mrm{\,eff}$. This is relatively straight-forward
for $+2$-particle states -- a Sudakov factor 
$\Delta(\facScale,\scale{2}^\mrm{\,eff})$ is applied to ensure a sensible 
result if the $\vec p_\perp$ of the combined Born$+2$-parton system is small.
For higher-multiplicity non-shower states, more scale hierarchies arise, and a 
more detailed scheme is necessary to cover all relevant cases. However, 
only two types of scale hierarchies can remain after removing the overlap 
between $n$-particle non-shower events and states that are produced by 
showering lower-multiplicity non-shower configurations: the ordering 
$\facScale > \scale{n}^\mrm{\,eff}$, or the ordering 
$\facScale > \scale{n-1}^\mrm{\,eff} > \scale{n}$\,\footnote{Consider a
non-shower (unordered) $+4$-particle state. After computing effective
scales, it is possible that a scale hierarchy 
$\facScale > \scale{2}^\mrm{\,eff} > \scale{3} > \scale{4}$ exists.
Such a configuration can be obtained in several ways showering 
lower-multiplicity non-shower states. \tit{a)} If the reconstructed underlying 
$+2$-particle state is not shower-like (i.e.\ unordered), then the $+4$-particle 
state with the above hierarchy can be produced by adding two ordered shower
emissions to the $+2$-particle state. Thus, the state is included by showering
a non-shower $+2$-particle state. \tit{b)}
If the reconstructed $+2$-particle state can be reached by an ordered sequence
of emissions, and furthermore $\scale{3} > \scale{4}$ then the ``unordering"
stems from the $+2$-particle to $+3$-particle transition. Thus, the 
$+4$-particle configuration can be reached by adding one ordered shower 
emission to a non-shower $+3$-particle state. In conclusion, the states with 
this more complex scale hierarchy should not be included through a non-shower 
$+4$-particle input, since this would result in over-counting.}. The hierarchy 
$\facScale > \scale{n}^\mrm{\,eff}$ is again ameliorated by applying a single
Sudakov factor $\Delta(\facScale,\scale{n}^\mrm{\,eff})$ to produce a sensible
result for small $\vec p_\perp$ of the combined Born$+n$-parton system. If instead
a hierarchy $\facScale > \scale{n-1}^\mrm{\,eff} > \scale{n}$ can be 
constructed, then a product of Sudakov factors 
$\Delta(\facScale,\scale{n-1}^\mrm{\,eff})
 \Delta(\scale{n-1}^\mrm{\,eff},\scale{n})$ is appropriate. This guarantees a 
uniform weighting of $+n$-particle events arising from either 
$+n$-particle non-shower states or showered $+(n-1)$-particle 
configurations. Note that the Sudakov factors $\Delta(\facScale,\scl^
\mrm{\,eff})$ are unity if $\facScale < \scl^\mrm{\,eff}$.

The information about the different types of scale hierarchies are also used to
remove the overlap between non-shower states with different parton 
multiplicities. States with scale hierarchies of the 
type $\scale{n-m}^\mrm{\,eff} > \scale{n-(m-1)} > \ldots > \scale{n}$ are 
removed for $m\ge 2$.
For states that contain the hierarchy $\scale{n-1}^
\mrm{\,eff} > \scale{n}$, the event is removed if the clustered 
$+(n-1)$-particle state is itself an unordered state.
Events without scale hierarchies that could have resulted from showering
lower-multiplicity states are kept; that includes all 
$+2$-particle states with unordered scales $\scale{2}>\scale{1}$ and 
$+1$-particle states with $\scale{1}>\facScale$.
For the interested reader we include further methodological instructions
in \appRef{app:overlap}.

\section{Results}
\label{sec:results}

In this section, we present results obtained with the new method, including
both the \mops factor and the non-shower states (called ``\mops + unordered" in the 
following). A detailed validation can be found in \appRef{app:validation}. 
The analyses are performed with \tsc{Rivet}~\cite{Buckley:2010ar}.
We begin this section by studying the effect of the new method on jet 
separations, before moving to comparisons to LHC data. In both cases, we 
juxtapose the results with the GKS \me corrections implemented in \Vincia. 
The GKS \mecs scheme includes emissions above the factorization scale
$\facScale$ (see \appRef{app:hardJets} for how those are generated) as does the 
\mops + unordered method by adding non-shower $+1$-particle states.
Emissions with scales $\scale{1}>\facScale$ would not naturally be present in
the pure or \mops corrected shower, where Born states are showered beginning
at $\facScale$. For the following results we add $+1$-particle 
states with scales $\scale{1}>\facScale$ explicitly to the pure and 
\mops corrected shower, and shower these states using $\scale{1}$ as shower 
starting scale. This decreases the significance of including non-shower states
w.r.t~comparing to a strictly ordered shower evolution, but should avoid 
using an ``overly conservative" shower setup when comparing to default \Vincia.

\subsection{Theory comparisons}
\label{sec:pheno}

\begin{figure}[p]
\centering
\begin{minipage}[b]{0.49\textwidth}
  \rivetFigureRatio{unordered-Zdecay/log10_y_mm+1.pdf}
  \rivetFigureRatio{unordered-Zdecay/log10_y_mm+1-ratio2.pdf}
  \rivetFigureRatio{unordered-Zdecay/log10_y_mm+1-ratio3.pdf}
  \rivetFigure{unordered-Zdecay/log10_y_mm+1-ratio4.pdf}
  \\
  \tbf{a)} $e^+e^-\to Z\to$~jets @ $91\,\mrm{Gev}$
\end{minipage}
\hfill
\begin{minipage}[b]{0.49\textwidth}
  \rivetFigureRatio{unordered-DY/log10_d_mm+1.pdf}
  \rivetFigureRatio{unordered-DY/log10_d_mm+1-ratio2.pdf}
  \rivetFigureRatio{unordered-DY/log10_d_mm+1-ratio3.pdf}
  \rivetFigure{unordered-DY/log10_d_mm+1-ratio4.pdf}
  \\
  \tbf{b)} $pp\to Z+$jets @ $7\,\mrm{TeV}$
\end{minipage}
\caption{\Pythia8.2.26\,+\,\Vincia2.2 predictions for jet resolution
measures $d_{m\,m+1}$ and $y_{m\,m+1}$ (the longitudinally 
invariant $k_\perp$ jet algorithm with $R=0.4$ for hadronic initial states and
the Durham jet algorithm for lepton collisions).
\me corrections are applied for $\le$ 3 emissions.
The red band is obtained by varying the effective scale 
$\scl^\mrm{\,eff}~[\mrm{GeV}]$ in non-shower events by factors of two.
\label{fig:unordUncertainties}}
\vspace*{7mm}
\centering
\begin{minipage}[b]{0.49\textwidth}
  \rivetFigure{results/3-runAS-HL/log10_d_mm+1.pdf}
\end{minipage}
\hfill
\begin{minipage}[b]{0.49\textwidth}
  \rivetFigureRatio{results/3-runAS-HL/log10_d_mm+1-ratio1.pdf}
  \rivetFigureRatio{results/3-runAS-HL/log10_d_mm+1-ratio2.pdf}
  \rivetFigureRatio{results/3-runAS-HL/log10_d_mm+1-ratio3.pdf}
  \rivetFigure{results/3-runAS-HL/log10_d_mm+1-ratio4.pdf}
\end{minipage}
\caption{\Pythia8.2.26\,+\,\Vincia2.2 and \Pythia8.2.15\,+\,\Vincia2.0.01 
predictions for jet resolution measures in Drell-Yan events  @ $7\,\mrm{TeV}$.
\me corrections are applied for $\le$ 3 emissions.
\label{fig:resultsZscales}}
\end{figure}

Here, the general features of the new method are illustrated by discussing
jet resolution scales. These variables show significant sensitivity to hard,
well-separated jets as well as parton-shower resummation, and they can thus be used 
to gauge the effect of different pieces in the calculation. To not obscure
the Sudakov shapes of the parton shower at low jet resolution, we do not include
multiparton interactions.

Hadron-level results for hadronic $Z$ decays and Drell-Yan events
are presented in \figRef{fig:unordUncertainties}.
The results have the expected behavior: at low jet resolution, parton-shower
effects dominate, while non-shower states contribute mainly to 
large jet scales. Hence, the \mops factor is dominating the observable at 
low scales. At LEP, shower states remain a dominant contribution even when 
modeling well-separated jets, and the effect of non-shower states remains
at below $10\%$ per bin. Results at the LHC are in stark contrast to
this. There, the influence of shower configurations decreases substantially for
large jet resolution, and non-shower phase-space regions become increasingly
important. The uncertainty from varying the effective scale 
is not significant at LEP, and should thus not be 
considered a realistic uncertainty estimate. At LHC, the variation of 
$\scl^\mrm{\,eff}~(=\facScale=\renScale=\startScale)$ is larger, and increases
for high jet resolution, as expected from varying scales in a tree-level 
fixed-order variation. At low 
resolution, we observe a small increase in the scale uncertainty, which stems
from the interplay of very large $\alpha_s$ values with the Sudakov factors
that are applied to non-shower states.

By comparing with previous ideas below, we hope to understand the 
short-comings and benefits of our \mops + unordered prescription.
In \figRef{fig:resultsZscales} we compare the results of \Vincia2.2 without
corrections, with the \mops correction, \mops + unordered scheme, and
\Vincia2.0.01 with smooth ordering for the GKS \mecd orders.

The \mops correction for purely evolution-induced events is small for all 
jet resolutions. Differences are mostly at the level of 
$1-5\%$, illustrating that the uncorrected shower already describes the matrix
elements well in phase-space regions reachable by showering.
As discussed above, the jet resolution scales exhibit a Sudakov suppression for
small values. In the Sudakov region, the corrected predictions should not 
deviate greatly from the ``plain" shower result. This is indeed the case for 
both the \mops + unordered and the GKS \mecs method for very small resolution
scales. The GKS \mecs method generates more events with larger $d_{m\,m+1}$ 
separation. Due to the unitarity of the shower, this leads to a depletion of 
events with small separation compared to the pure shower. 
The behavior is consistent with the findings in~\cite{Fischer:2016vfv}, where 
differences between strong and smooth ordering have been investigated.
The impact of non-shower states in the \mops + unordered scheme remains 
noticeable close to the peak of the distribution, although the modeling of 
the Sudakov region approaches the uncorrected shower more quickly
than for the GKS \mecs method. This means that the handling of non-shower states
with large scale hierarchies (cf.~end of~\secRef{sec:summary}) is important in
our approach. Merging approaches commonly discard non-shower states with 
separation below a certain (merging) scale.

In conclusion, we believe that the \mops + unordered scheme 
has desirable features, and that the choices in the method lead to the 
expected behavior.  

\subsection{Comparisons to data}
\label{sec:compToDatas}

\begin{figure}[t]
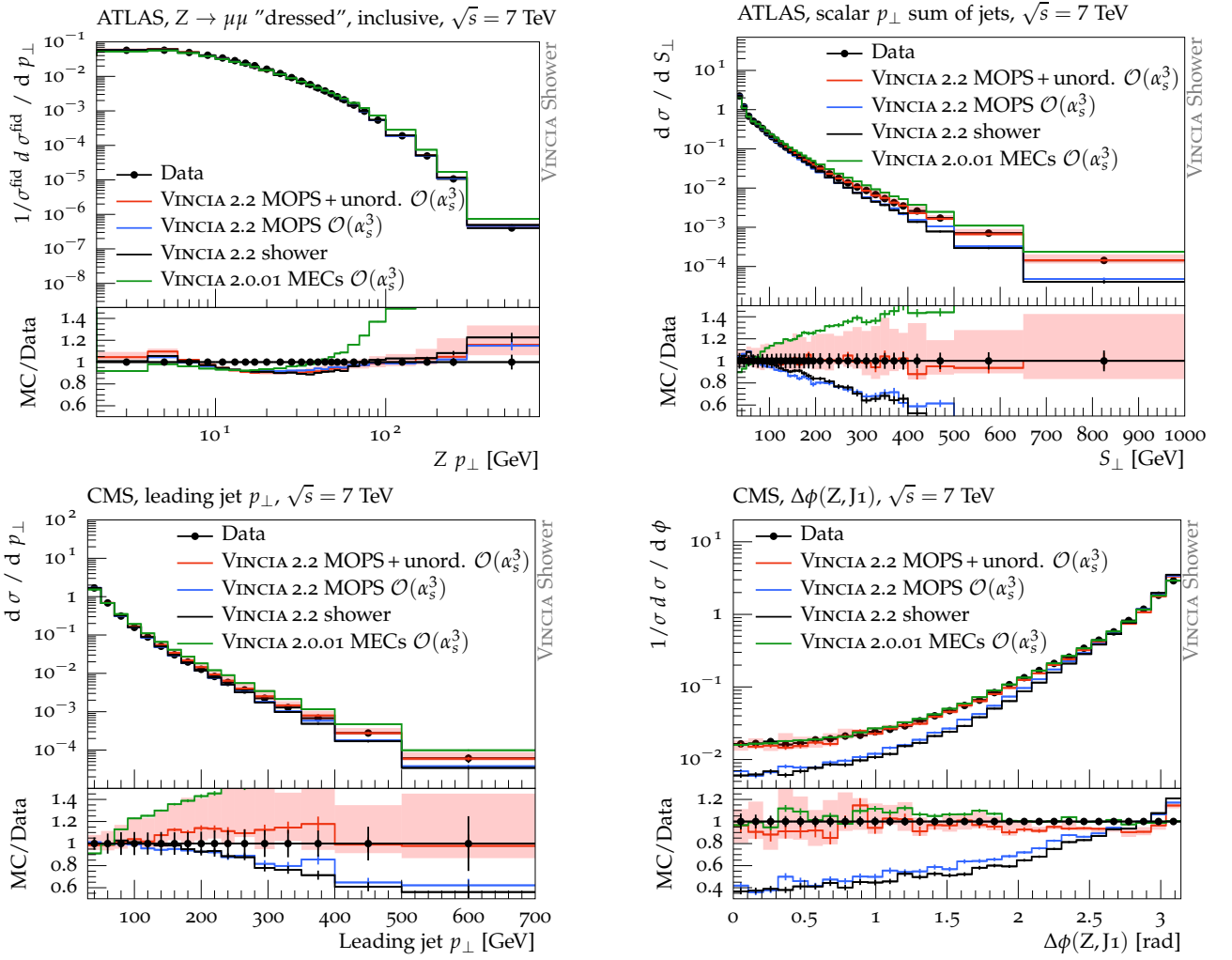

\centering
\begin{minipage}[b]{0.49\textwidth}
  \rivetFigure{results/3-runAS-HL/ZpT.pdf}
  \rivetFigure{results/3-runAS-HL/JpT.pdf}
\end{minipage}
\hfill
\begin{minipage}[b]{0.49\textwidth}
  \rivetFigure{results/3-runAS-HL/ST.pdf}
  \rivetFigure{results/3-runAS-HL/deltaPhiZJ.pdf}
\end{minipage}
\caption{\Pythia8.2.26\,+\,\Vincia2.2 and \Pythia8.2.15\,+\,\Vincia2.0.01 
predictions compared to ATLAS data from \cite{Aad:2013ysa,*Aad:2014xaa} and
CMS data from \cite{Chatrchyan:2013tna,*Khachatryan:2014zya}. \tsc{Rivet} 
analyses ATLAS$\_\,$2013$\_\,$I1230812, ATLAS$\_\,$2014$\_\,$I1300647, 
CMS$\_\,$2013$\_\,$I1209721, and CMS$\_\,$2015$\_\,$I1310737. 
For the leading jet $p_\perp$ and the scalar $p_\perp$ sum of jets the
predictions are rescaled to the experimental inclusive one-jet cross section.
\me corrections are applied for $\le$ 3 emissions.
The red band is obtained by varying the effective scale 
$\scl^\mrm{\,eff}~[\mrm{GeV}]$ in non-shower events by factors of two.
\label{fig:resultsZdata}}
\end{figure}

To assess how the method performs for realistic observables, we now turn 
to Drell-Yan + jets measurements at the LHC. All curves employ the NNPDF\,2.1 
LO PDF set~\cite{Ball:2011uy} and use the corresponding strong coupling
$\as(k_\mu\,t)$ with one-loop running, $\as(m_Z^2)=0.13$, and $k_\mu=1$ 
for all branchings. We use these settings to compare all schemes on equal
footing and choose $k_\mu=1$ as required for the calculation of the effective
scale\,\footnote{Different $k_\mu$ values for different branching types
invalidate the interpretation of the effective scale as a single parton-shower 
starting scale for subsequent showering.}.
Soft-physics parameters are kept at their current \Vincia default
values. The default \Vincia2.0.01 tune~\cite{Fischer:2016vfv} corresponds to 
different $\as$ settings. While this results in a slightly better data
description, it does not alter the general observations and conclusions of this
section.

In \figRef{fig:resultsZdata} we confront the results of \Vincia2.2 without
corrections, with the \mops correction, \mops + unordered scheme, and
\Vincia2.0.01 with GKS \me corrections with ATLAS~\cite{Aad:2013ysa,*Aad:2014xaa}
and CMS~\cite{Chatrchyan:2013tna,*Khachatryan:2014zya} measurements. 

As already seen in \secRef{sec:pheno}, the effect of the \mops correction
factor is small for all observables. This is of benefit for the description
of the Drell-Yan $p_\perp$ spectrum (upper left panel of 
\figRef{fig:resultsZdata}), for which the plain shower already offers a sensible
data description. The quality of the description also remains unchanged for 
the \mops + unordered scheme. The other observables in \figRef{fig:resultsZdata}
test the existence of hard, well-separated emissions in the tails of the
distributions and are thus poorly modeled with the parton shower alone. We find
a very good data description with the \mops + unordered scheme. In particular,
the quality of the data description in our scheme relies crucially on the
treatment of non-shower states. The scale-setting mechanism presented in
\secRef{sec:nonShower} produces promising results, with the naive central
scale choice close to the data,
but with a large, leading-order-like uncertainty due to scale variations. We
anticipate that the width of the band will decrease when performing a 
next-to-leading-order calculation with a similar scale choice. The uncertainty
due to scale variations is largest in phase-space regions most sensitive to
non-shower contributions. For the $S_\perp$ and leading jet $p_\perp$ 
distributions, the results of the GKS \mecs approach touch the uncertainty bands
attributed to non-shower events at low values, but are outside of the band in
regions influenced by multiple hard jets. Both of these observables are much
improved in the \mops + unordered method, compared to the uncorrected shower.
For the angle between the $Z$-boson and the hardest jet we observe a satisfactory 
data description for both our new method and \Vincia2.0.01.

\section{Conclusions}
\label{sec:conclusions}

We have presented an algorithm to obtain fixed-order accurate predictions
combined with all-order parton-shower evolution that produces finite and
non-overlapping results without introducing a merging scale. The new
algorithm requires the introduction of a sophisticated matrix-element
correction method for 
evolution-induced configurations. States beyond the reach of the parton shower
are included with a systematic scale-setting procedure. This smoothly combines
non-shower configurations and states produced in the ordered parton-shower
evolution. The algorithm does not depend on specific properties of the
parton shower and allows for arbitrary dead zones (which may be required by
resummation considerations). The new fixed-order + parton-shower scheme has 
been implemented in the \Vincia
parton shower and will be made publicly available upon the \Vincia2.2
release.

The effect of including \me corrections for ordered parton-shower splittings
is minor compared to the uncorrected shower. This means that the method does not
deteriorate the shower resummation, and gives us confidence that the 
improvement does not interfere with other improvement strategies 
\cite{Li:2016yez,*Hoche:2017iem,*Hoche:2017hno}. The 
main improvements stem from a careful treatment of contributions from 
phase-space regions that are not accessible by ordered parton showers. 
Such contributions are included with a sophisticated scale-setting prescription.
For hadronic initial we find the scale setting to have a sizable influence on 
observables, since large parts of phase space are not shower accessible. 
We presented comparisons to data for the $pp\to Z+$jets process and 
found the results of our new algorithm to be in good agreement
with the data.

\section*{Acknowledgements}

We would like to thank Stefan H\"{o}che and Peter Skands for helpful 
discussions, and Peter Skands for comments on the manuscript.
This work was supported by the Australian Research Council and by Fermi
Research Alliance, LLC under Contract No. DE-AC02-07CH11359 with the U.S.
Department of Energy, Office of Science, Office of High Energy Physics.
NF thanks the SLAC and Fermilab theory divisions
for hospitality during the course of this work.

\appendix

\section{Review of GKS matrix-element corrections}
\label{app:GKSMECs}

Iterative \me corrections have first been introduced in \cite{Giele:2011cb},
and have been applied to colorless resonance decays~\cite{Giele:2011cb} as well
as to initial-state radiation~\cite{Fischer:2016vfv}.
Finite multiplicative correction factors are applied order by 
order in perturbation theory as the shower evolves. The \mec factor 
$\MEC{\pPhi{n+1}}$ replaces the splitting kernels by a ratio of tree-level 
matrix elements. Symbolically, the correction factor can be written as
\begin{align}
  P({\scriptstyle{\pPhi{n+1}/\pPhi{n}}})~~\longrightarrow~~
  \MEC{\pPhi{n+1}}~P({\scriptstyle{\pPhi{n+1}/\pPhi{n}}})\equiv
  \frac{\ME{\pPhi{n+1}}}{\sum\limits_{\pPhi{n}'}
  P({\scriptstyle{\pPhi{n+1}/\pPhi{n}'}})~
  \ME{\pPhi{n}'}}~P({\scriptstyle{\pPhi{n+1}/\pPhi{n}}})~.
  \label{eq:GKSmecFactor1}
\end{align}
The denominator sums over all possible $n$-particle states 
through which the shower could have produced the $(n+1)$-particle state.

\subsection{Smoothly ordered showers}
\label{app:smoothMarkov}

The \mec formalism in~\cite{Giele:2011cb,Fischer:2016vfv} requires a 
history-independent parton shower that covers the full phase space for the 
\mecd orders. Therefore, \Vincia introduces the concept of smooth ordering. At 
any stage of the evolution the following procedure determines at which scale 
the shower off each parton in a $(n+1)$-particle state is restarted:
\begin{newItem}
\item Find all physical clusterings $\pPhi{n+1}\to \pPhi{n}^i$ and their
branching scales $\,t({\scriptstyle{\pPhi{n+1}/\pPhi{n}^i}})$. The 
reference scale is the minimum of all scales, $\hat t({\scriptstyle{\pPhi{
n+1}}})=\mrm{min}_{\,i}\left(t({\scriptstyle{\pPhi{n+1}/\pPhi{n}^i}})\right)$. 
\item Divide the $(n+1)$-particle state into a set of ``ordered" and ``unordered 
partons". For more details see~\cite{Fischer:2016vfv}.
\item The evolution of ``ordered partons" is restart at the reference scale $\hat t$.
``Unordered partons" are allowed to radiate up to the phase-space maximum, but with 
the suppression factor 
\begin{align}
  \pimp\left(\hat t({\scriptstyle{\pPhi{n+1}}}),t({\scriptstyle{\pPhi{n+2}/
  \pPhi{n+1}}})\right) = \frac{\hat t({\scriptstyle{\pPhi{n+1}}})}
  {\hat t({\scriptstyle{\pPhi{n+1}}}) +
  t({\scriptstyle{\pPhi{n+2}/\pPhi{n+1}}})}~.
  \label{eq:Pimp}
\end{align}
\end{newItem}
When taking smooth ordering into account, the \mec factor 
\eqref{eq:GKSmecFactor1} should be defined as
\begin{align}
  \MEC{\pPhi{n+1}} = \frac{\ME{\pPhi{n+1}}}{\sum\limits_{\pPhi{n}'}
  \mc O\left(\hat t({\scriptstyle{\pPhi{n}'}}),
  t({\scriptstyle{\pPhi{n+1}/\pPhi{n}'}})\right)~
  P({\scriptstyle{\pPhi{n+1}/\pPhi{n}'}})~\ME{\pPhi{n}'}}~.
\end{align}
The ordering criterion reflects the different treatment of partons,
\begin{align}
  \mc O\left(\hat t({\scriptstyle{\pPhi{n}'}}),
  t({\scriptstyle{\pPhi{n+1}/\pPhi{n}'}})\right) = \left\{\begin{matrix} 
  \pimp\left(\hat t({\scriptstyle{\pPhi{n}'}}),
  t({\scriptstyle{\pPhi{n+1}/\pPhi{n}'}})\right) & 
  \text{for a branching of an ``unordered parton",} \\[3mm] 
  \Theta\left(\hat t({\scriptstyle{\pPhi{n}'}}) - t({\scriptstyle{\pPhi{n+1}/
  \pPhi{n}'}})\right) & \text{for a branching of an ``ordered parton".}~~\,\,
  \end{matrix}\right.
  \nonumber
\end{align}
The procedure guarantees a history-independent parton shower that covers
the full kinematic range. However, it introduces complications that are hard
to constrain from QCD considerations alone.

\subsubsection{Sudakov factors in unordered regions}
Consider the exclusive Born+jet cross section at the end of parton shower with
the following evolution. The shower starts at the factorization scale of
the Born process $\facScale$. After the branching at scale $t_1<\facScale$,
all partons explore their full kinematic range up to the scale $\maxScale$ 
and are evolved down to the  shower cut-off $\cutoff$. Dropping the PDF factor 
for the second leg and suppressing most dependences of the splittings kernels,
the exclusive cross section for this evolution sequence reads
\begin{align}
  \dSig{1}{\cutoff} = \noEmi{1}{\maxScale}{\cutoff}~\cdot~
  \alpha_s(\scale{1})~P(\scale{1})~\dfrac{\PDF{1}{x_1}{\scale{1}}}{\PDF{0}{x_0}
  {\scale{1}}}~\noEmi{0}{\facScale}{\scale{1}}~\cdot~\PDF{0}{x_0}{\facScale}~
  \ME{\Phi_0}~\dPhi{1}~.
\end{align}
The no-emission probability $\noEmi{1}{\maxScale}{\cutoff}$ can be split up 
into an ordered part $\noEmi{1}{\scale{1}}{\cutoff}$ and a part that 
reflects the evolution in the unordered region $\noEmiUO{1}{\maxScale}
{\scale{1}}$. We use the relation~\cite{Ellis:1991qj}
\begin{align}
  \noEmi{n}{\scale{n}}{\scale{n+1}} = \frac{\PDF{n}{x_n}{\scale{n+1}}}
  {\PDF{n}{x_n}{\scale{n}}}~\sudakov{n}{\scale{n}}{\scale{n+1}}
  \label{eq:PiDelta}
\end{align}
to write the cross section in terms of Sudakov factors,
\begin{align}
  \dSig{1}{\cutoff} = \noEmiUO{1}{\maxScale}{\scale{1}}~
  \cdot~\PDF{1}{x_1}{\cutoff}~\sudakov{1}{\scale{1}}{\cutoff}~  
  \alpha_s(\scale{1})~P(\scale{1})~
  \sudakov{0}{\facScale}{\scale{1}}~\cdot~\ME{\Phi_0}~\dPhi{1}~.
\end{align}
The no-emission probability $\noEmiUO{1}{\maxScale}{\scale{1}}$ remains in the 
cross section. In \Vincia this factor is defined as
\begin{align}
  \noEmiUO{1}{\maxScale}{\scale{1}} = 
  \exp\left(-\sum_{1~\to~2}\int\d z~
  \int_{\scale{1}}^{\maxScale}\d\scl~
  \frac{\PDF{2}{x_2}{\scale{1}}}{\PDF{1}{x_1}{\scale{1}}}~
  \alpha_s(t)~\pimp~P(t,z)\right)~.
  \label{eq:noEmiUO}
\end{align}
Here, the scale in the PDF ratio is fixed to the scale of the previous 
emission to ensure the proper cancellation between PDF factors for branchings 
in the unordered region. However, \eqref{eq:noEmiUO} does not have a direct
correspondence to any term in the DGLAP equation reformulated as a backwards
evolution~\cite{Sjostrand:1985xi}.

\subsubsection{Missing evolution and configurations}
For low multiplicities, all partons in the system are treated as unordered and 
explore their phase space up to the kinematics limit. However, starting for 
higher multiplicities, ``ordered partons" are present which restart their 
evolution at the Markovian scale. 
By definition, this scale is smaller or equal to the scale of the last 
branching. The allowed branching range of ``ordered partons" is therefore
more restricted than in an ordered shower.

As with every parton shower that only contains QCD splittings, certain flavor 
configurations cannot be 
reached, independent of kinematic constraints. One such example is $q\bar q\to 
W q\,'\bar q\,''$, where the $W$ boson can only be radiated off the final-state 
legs. To include such a configuration within the \mecs method an electroweak 
shower is necessary.

\subsection{The treatment of hard jets}
\label{app:hardJets}

To avoid the concept of ``power showers'' and simultaneously allow jets with 
scales $\scl>\facScale$, \Vincia distinguishes between non-QCD and QCD processes. 
The latter category covers all hard processes with partons in the final state
(except partons arising from resonance decay).

\begin{figure}[t]
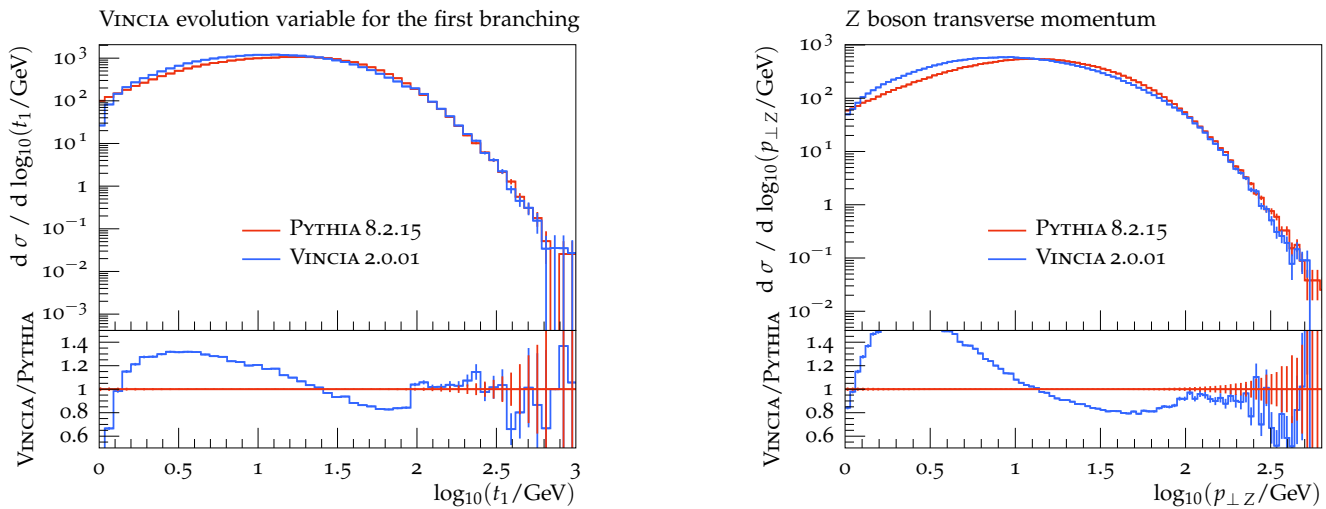

\centering
\rivetFigure{eventSamples/20-log10_TMS_1.pdf}
\hfill
\rivetFigure{eventSamples/01-log10_ZpT.pdf}
\caption{Distribution of the \Vincia evolution variable after the first branching 
(\tit{left}) and the $Z$ boson transverse momentum (\tit{right}) for 
$pp\to Z+$jets at parton level. 
\label{fig:DrellYanT1}}
\end{figure}

In non-QCD processes the input events are divided in two samples. The first one 
is associated with no hard jets, while the second sample contains at least one 
jet with $\scl>\facScale$. Because both samples are weighted differently, this 
introduces a non-smooth transition, see the left panel of \figRef{fig:DrellYanT1}.
When more branchings are taken into account, the effect is washed out and the 
step barely visible as shown in the right panel of \figRef{fig:DrellYanT1}.

The first emission off a QCD $2\to2$ process is treated similar to the 
procedure summarized in \appRef{app:smoothMarkov}: all partons are 
allowed to explore their full phase space, but with a suppression of
\begin{align}
  \pimp\left(\facScale,t({\scriptstyle{\pPhi{1}/
  \pPhi{0}}})\right) = \frac{\facScale}{\facScale +
  t({\scriptstyle{\pPhi{1}/\pPhi{0}}})}~.
\end{align}
Here the factorization scale replaces the Markovian reference scale. This 
leads to similar, leftover no-emission probabilities from unordered regions
as discussed in \appRef{app:smoothMarkov}.

\section{Validation of matrix-element corrections for ordered emissions
in \Vincia}
\label{app:validation}

In this section we validate the numerical implementation of the \mops method 
in \Vincia by comparing it to merged predictions using the CKKW-L merging 
implementation in \Pythia8~\cite{Lonnblad:2011xx} applied
to \Vincia. For the latter we define the merging scale as the minimum of
all evolution scales, $\scale{\t{MS}}=\mrm{min}_{\,i}\left(
t({\scriptstyle{\pPhi{n+1}/\pPhi{n}^i}})\right)$.
No color information is used to find possible clusterings.
For the validation we use parton-level results with a fixed $\as$ for both 
methods and do not include events that cannot be reproduced by \Vincia with 
an ordered sequence of branching scales. To ensure the same Sudakov factors 
\me corrections are also applied in the case of merging.

\FigsRef{fig:valFSR1} and \ref{fig:valISR1} show a comparison between the 
results of the \mops method and merging including a \mecd first emission. Each 
simulation contains at least $10^8$ 
input events generated with \Madgraph~\cite{Alwall:2007st,*Alwall:2011uj}. 
The lower panels present the deviation between the two methods, normalized to 
the statistical uncertainty of the merged prediction in the respective bin. 
As both methods should provide the same result, this distribution should
exhibit statistical fluctuations only.
Parton-level results for $e^+e^-\to Z\to$~jets and $\tau^+\tau^-
\to H\to$~gluons are presented in \figRef{fig:valFSR1}. The deviation in the
lower panels clearly show that both methods are identical up to 
statistical fluctuations.
Similar plots are shown for on-shell $Z$-boson production in 
\figRef{fig:valISR1}. 
Note that we exclude branchings with scales above the factorization scale for 
comparison purposes. This is necessary due to how such emissions are generated in 
\Vincia, see \appRef{app:hardJets}.

When correcting the second emission, we expect slight mismatches
between the predictions of the two methods. The matrix elements in \Vincia 
are taken from \Madgraph4. It would thus be preferable to use \Madgraph4 input for 
the merging. However, \Madgraph4 is no longer developed and does not allow for
linking against \lhapdf5~\cite{Whalley:2005nh}, while \Pythia8 requires
\lhapdf5 or higher. Thus, using the same PDF set for hadronic initial states
means that the input for merging was generated with \Madgraph5.
\Madgraph4 and 5 exhibit shape and normalization differences at the 
(sub-)percent level in the observables investigated for the validation,
as discussed in the following. As an example, we compare the \me output 
of \Madgraph4 and 5 for $e^+e^-\to Z\to q\bar qgg$ with a cut on the invariant 
mass of jet pairs, $m_{jj}\ge 5~\mrm{GeV}$.
We further include curves for the \Vincia matrix element integrated with 
\tsc{Rambo}~\cite{Kleiss:1985gy} (an implementation of which is included in 
\Vincia) and normalized to the \Madgraph4 cross section, as we are mainly 
interested in shape differences. 
The results are shown in \figRef{fig:valMG}. The ratio plots shown in the lower
panels reveal differences between all three predictions, mostly at the level of 
around $0.5\%$. While those mismatches are irrelevant in practical studies, 
they deteriorate the quality the validation. Nevertheless the results of the 
validation are satisfactory.
When correcting the third emission, we anticipate further 
differences between the two methods.
In \Vincia, the color matrices for matrix elements with two identical quark
pairs and at least one gluon are decomposed by hand; see \cite{Fischer:2016vfv}.
Therefore, higher orders cannot be validated at the same level as the first 
order.

In \figRef{fig:val3} we show a comparison of merging and the \mops method for
three corrected emissions. The lower panels show the ratio of predictions 
with the \mops method to merged results. Small deviations between the two 
methods are visible at large scales. Considering that the differences 
are at most $3\%$, and that we expect some mismatches, and that the differences
are mostly in a region where non-shower states have a very large impact 
(cf.~\figRef{fig:resultsZscales}), we find the methods in good agreement.

\begin{figure}[p]
\centering
\begin{minipage}[t]{0.49\textwidth}
  \rivetFigureRatio{validation-Zdecay/1-fixAS-strong-FS/log10_y_mm+1.pdf}
  \rivetFigureRatio{validation-Zdecay/1-fixAS-strong-FS/log10_y_mm+1-ratio1.pdf}
  \rivetFigure{validation-Zdecay/1-fixAS-strong-FS/log10_y_mm+1-ratio2.pdf}
  \\[-1mm]
  \rivetFigure{validation-Zdecay/1-fixAS-strong-FS/20-log10_TMS_3.pdf}
  \\
  \tbf{a)} $e^+e^-\to Z\to$~jets
\end{minipage}
\hfill
\begin{minipage}[t]{0.49\textwidth}
  \rivetFigureRatio{validation-Hdecay/1-fixAS-strong-FS/log10_y_mm+1.pdf}
  \rivetFigureRatio{validation-Hdecay/1-fixAS-strong-FS/log10_y_mm+1-ratio1.pdf}
  \rivetFigure{validation-Hdecay/1-fixAS-strong-FS/log10_y_mm+1-ratio2.pdf}
  \\[-1mm]
  \rivetFigure{validation-Hdecay/1-fixAS-strong-FS/20-log10_TMS_3.pdf}
  \\
  \tbf{b)} $\tau^+\tau^-\to H\to$~gluons
\end{minipage}
\caption{Parton-level results: the distribution of the merging scale in
exclusive 3-parton events (\tit{bottom}) and the logarithmic distributions of 
differential jet resolutions (\tit{top}).
Merged predictions with a merging-scale value of $5~\mrm{GeV}$ are
compared to predictions with the \mops method.
\label{fig:valFSR1}}
\end{figure}

\begin{figure}[p]
\centering
\begin{minipage}[t]{0.48\textwidth}
  \rivetFigureRatio{validation-DY/1-fixAS-strong-FS/log10_d_mm+1.pdf}
  \rivetFigureRatio{validation-DY/1-fixAS-strong-FS/log10_d_mm+1-ratio1.pdf}
  \rivetFigure{validation-DY/1-fixAS-strong-FS/log10_d_mm+1-ratio2.pdf} \\
  \rivetFigure{validation-DY/1-fixAS-strong-FS/20-log10_TMS_1.pdf}
  \caption{Parton-level results for $pp\to Z+$jets: the distribution
  of the merging scale in exclusive 1-parton events (\tit{bottom})
  and the logarithmic distributions of differential jet resolutions (\tit{top}). 
  Merged predictions with a merging-scale value of $5~\mrm{GeV}$ are
  compared to predictions with the \mops method. \label{fig:valISR1}}
\end{minipage}
\hfill
\begin{minipage}[t]{0.48\textwidth}
  \rivetFigure{validation-Zdecay/ZtoQGGQB/00-log10_y_23.pdf} \\[1mm]
  \rivetFigure{validation-Zdecay/ZtoQGGQB/01-log10_y_34.pdf} \\
  \rivetFigure{validation-Zdecay/ZtoQGGQB/21-log10_TMS_4.pdf}
  \caption{Parton-level results for $e^+e^-\to Z\to q\bar qgg$: the
  distribution of the merging scale in exclusive 4-parton events 
  (\tit{bottom}) and the logarithmic distributions of differential jet 
  resolutions. Comparison of \Madgraph4, \Madgraph5, and 
  \Vincia+\,\Madgraph4\,+\,Rambo. \label{fig:valMG}}
\end{minipage}
\end{figure}

\begin{figure}[p]
\centering
\begin{minipage}[t]{0.49\textwidth}
  \rivetFigureRatio{validation-Zdecay/3-fixAS-strong-FS/log10_y_mm+1.pdf}
  \rivetFigureRatio{validation-Zdecay/3-fixAS-strong-FS/log10_y_mm+1-ratio1.pdf}
  \rivetFigure{validation-Zdecay/3-fixAS-strong-FS/log10_y_mm+1-ratio2.pdf}
  \\[-1mm]
  \rivetFigureRatio{validation-Zdecay/3-fixAS-strong-FS/log10_TMS_m.pdf}
  \rivetFigureRatio{validation-Zdecay/3-fixAS-strong-FS/log10_TMS_m-ratio1.pdf}
  \rivetFigure{validation-Zdecay/3-fixAS-strong-FS/log10_TMS_m-ratio2.pdf}
  \\
  \tbf{a)} $e^+e^-\to Z\to$~jets
\end{minipage}
\hfill
\begin{minipage}[t]{0.49\textwidth}
  \rivetFigureRatio{validation-DY/3-fixAS-strong-FS/log10_d_mm+1.pdf}
  \rivetFigureRatio{validation-DY/3-fixAS-strong-FS/log10_d_mm+1-ratio1.pdf}
  \rivetFigure{validation-DY/3-fixAS-strong-FS/log10_d_mm+1-ratio2.pdf}
  \\[-1mm]
  \rivetFigureRatio{validation-DY/3-fixAS-strong-FS/log10_TMS_m.pdf}
  \rivetFigureRatio{validation-DY/3-fixAS-strong-FS/log10_TMS_m-ratio1.pdf}
  \rivetFigure{validation-DY/3-fixAS-strong-FS/log10_TMS_m-ratio2.pdf}
  \\
  \tbf{b)} $pp\to Z+$jets
\end{minipage}
\caption{Parton-level results: the distribution of the merging scale in
exclusive 4- and 5-parton events (\tit{bottom}) and the logarithmic 
distributions of differential jet resolutions (\tit{top}).
Merged predictions with a merging-scale value of $5~\mrm{GeV}$ are
compared to predictions with the \mops method.
\label{fig:val3}}
\end{figure}

\section{Identifying and removing the overlap between states with different 
multiplicities}
\label{app:overlap}

As discussed in \secsRef{sec:nonShower} and \ref{sec:summary}, overlap
between (the shower off) non-shower states with different parton
multiplicities exists and has to be removed. In this section we briefly
explain, for interested readers and practitioners, how different states are 
treated to remove potential overlap.

\textbf{$\mathbf{+0}$-particle states:} The shower is started at the 
factorization scale $\facScale$ of the Born state and no further restrictions 
apply.

\textbf{$\mathbf{+1}$-particle states:} Only events where all scales $t_1$ 
exceed the factorization scale, $\scale{1}>\facScale$, are taken into account. 
After a path is chosen, the shower off the $+1$-particle state starts at the 
scale $t_1$.

\textbf{$\mathbf{+2}$-particle states:} To avoid overlap with the shower off 
non-shower $+1$-particle states, an ordering of the clustering scales with 
respect to the factorization scale is not checked. Only events, where $\scale{2}
>\scale{1}$ holds for all paths, are taken into account and the effective
scale $\scale{2}^{\,\mrm{eff}}$ is calculated. If $\facScale>\scale{2}^
{\,\mrm{eff}}$ a Sudakov factor $\Delta(\facScale,\scale{2}^{\,\mrm{eff}})$ is 
attached by trial-showering the clustered Born state. The shower off the 
$+2$-particle state starts at $\scale{2}^{\,\mrm{eff}}$.

\textbf{$\mathbf{+n}$-particle states $(n\ge3)$:} As for the non-shower 
$+2$-particle 
states, an ordering of the clustering scales with respect to the factorization 
scale is not checked. Only events without an ordered path are taken into 
account. The effective scales $\scale{2}^{\,\mrm{eff}},~\scale{3}^
{\,\mrm{eff}},~\ldots~\scale{n}^{\,\mrm{eff}}$ are calculated
and the smallest $k\in\{2\ldots n\}$ which leads to an ordered sequence of scales,
$\scale{k}^\mrm{\,eff} > \scale{k+1} > \ldots > \scale{n}$, is found. If
$k\le n-2$, the event is removed from consideration due to overlap with
showering lower-multiplicity non-shower states, see \secRef{sec:summary}.
If $k=n-1$, i.e. $\scale{n-1}^\mrm{\,eff} > \scale{n}$, the event is
removed, if the clustered $+(n-1)$-particle state is itself a non-shower
state. 
For events that are not rejected we chose one of the paths for which 
$\scale{n-1}^\mrm{\,eff} > \scale{n}$ holds and attached the Sudakov
factors $\Delta(\facScale,\scale{n-1}^{\,\mrm{eff}})\Delta(\scale{n-1}^
{\,\mrm{eff}},\scale{n})$. The shower off the $+n$-particle state
starts at $\scale{n}$. If no scale hierarchy is found, the event is retained,
the Sudakov factor $\Delta(\facScale,\scale{n}^{\,\mrm{eff}})$ is
attached, and the $+n$-particle states is showered from $\scale{n}^
{\,\mrm{eff}}$.

\end{document}